\documentclass[slac_one]{revtex4}
\usepackage{graphicx}
\usepackage{subfigure}
\usepackage{fancyhdr}
\usepackage{SIunits}
\usepackage{amsmath}
\bibpunct{[}{]}{,}{n}{}{;}% 
\begin{document}
%
%Title of paper
\title{Jet production cross section and jet properties
in $pp$ collisions at $\sqrt{s} = 7~\tera\electronvolt$
with the ATLAS detector} %% Paper title goes here
\author{Zinonos Zinonas}
\affiliation{Universit\`{a} di Pisa \& INFN Sezione di Pisa, Italia}
\begin{abstract}
Jet cross sections have been measured with the ATLAS detector for the first time in
proton-proton collisions at center-of-mass energy of $7~\tera\electronvolt$. 
The measurement uses an integrated luminosity of $17~\nano\barn^{-1}$ recorded at the Large Hadron Collider. 
The anti-$k_t$ algorithm is used to identify jets, with jet resolution parameters $R=0.6$. 
The dominant uncertainty comes from the jet energy scale, which is determined to within $7\%$ for central jets above $60~\giga\electronvolt$ 
transverse momentum. 
Inclusive single-jet differential cross sections are presented as functions of jet transverse momentum and rapidity. 
Dijet cross sections are presented as functions of dijet mass and angle $\chi$. 
The experimental results are compared to the expectations based on next-to-leading order QCD.
\end{abstract}
\maketitle
\thispagestyle{fancy}
\section{INTRODUCTION}

In proton-proton ($pp$) collisions produced by the Large Hadron Collider (LHC) at high center-of-mass energy ($\sqrt{s}$), the production of jets is 
the dominant high transverse momentum ($p_T$) process.

Inclusive jet and dijet cross sections have been measured with the ATLAS detector \cite{ATLAS-CONF-2010-050}
for first time at an energy more than 3 times higher than that of the measurements done at Tevatron. 
The measurements presented here are performed using the data collected between 30 March and 5 June 2010, when the LHC
delivered $pp$ collisions at $\sqrt{s} = 7~\tera\electronvolt$. 
The data correspond to an integrated luminosity of $17\pm 2~\nano\barn^{-1}$ and have negligible impact from pile-up events. 
The integrated luminosities are calculated run-by-run using the Van der Meer calibration technique \cite{ATLAS-CONF-2010-050}.

The measured cross sections extend up to a jet $p_T$ of $600~\giga\electronvolt/c$ and dijet masses of $2~\tera\electronvolt/c^2$.
Leading logarithmic (LO) parton-shower Monte Carlo (MC) programs provide a reasonable description of the shapes of the measured cross section
distributions and of the energy flow in the proximity of the jets. All measurements are well described by next-to-leading order (NLO) 
perturbative calculations corrected for non-perturbative QCD effects.
\section{ANALYSIS}
% \paragraph{Cross section definitions} 
Jets are identified using the infrared-safe anti-$k_T$ algorithm \cite{ATLAS-CONF-2010-050}
with size parameter $R=0.6$, and using three-dimensional calorimeter clusters as input with full $4$-momenta recombination.
Jet energies are calibrated using a MC-derived $p_T$ and $\eta$-dependent calibration scheme \cite{ATLAS-CONF-2010-050}.
Afterwards, jet $4$-momenta are corrected for the primary vertex position and for detector effects back to the hadron level.

Inclusive single-jet double-differential cross sections are measured as a function of the jet $p_T$ in slices of absolute rapidity ($|y|$).
Dijet cross sections are measured as a function of the dijet mass in bins of 
$|y|_\text{max} = \text{max} (|y_1 |,\, |y_2 |)$ and of the 
dijet angle $\chi =  \exp{|y_1 -y_2|} \simeq \tfrac{1+\cos\theta^\ast}{1-\cos\theta^\ast}$, where $\theta^\ast$ 
is the polar scattering angle in the dijet center-of-mass frame.

It is also important that the energy flow around the jet cone is well understood and described by MC models. 
The energy and momentum flow within jets can be expressed in terms of the differential jet shape defined as the fraction, $\rho$, 
of the jet momentum contained within a ring of thickness $\Delta r = 0.1$ at a radius $r = \sqrt{ (\Delta\eta)^2 + (\Delta\phi )^2 }$ 
around the jet axis, divided by $\Delta r$. This variable offers a useful indication of the sensitivity of the reconstructed jets to the hadron 
fragmentation and underlying event.    

\texttt{Pythia 6.4} event generator, 
with \texttt{MSTW L0}$\ast$ parton density function and 
a set of parameters (\texttt{ATLAS MC09}) tuned to describe the exisiting minimum bias and underlying event data \cite{ATLAS-CONF-2010-050}, 
was used to simulate jets in $pp$ collisions at $\sqrt{s} = 7~\tera\electronvolt$. These samples are used for the baseline comparisons, 
the study of the systematic uncertainties and for data corrections.
%
%
%
% \paragraph{Ingredients of the data analysis } 
The jet algorithm is run on calorimeter clusters assuming that the event vertex is at the origin. 
The jet $4$-momentum is then recalculated with respect to the $z$-position of the primary vertex. 
After calibration, all events are required to have at least one jet within the kinematic region 
$p_T > 60~\giga\electronvolt/c$, $|y| < 2.8$. Any subleading jets should have  $p_T > 30~\giga\electronvolt/c$.
Additional quality criteria are also applied to ensure that jets are not produced by single noisy calorimeter 
cells nor problematic detector regions \cite{ATLAS-CONF-2010-050}.
Events are required to have at least one vertex consistent with the beamspot position, 
with at least five tracks connected to it, and with $|z| < 10~\centi\meter$. 
The $z$ vertex distribution in the MC simulation is reweighted to agree with that in data. 
Background contributions from non-$pp$-collision sources were evaluated using unpaired and
empty bunches and found to be negligible \cite{ATLAS-CONF-2010-050}.

Two types of triggers are used to select events in this analysis.
A minimum bias  trigger,  which requires a single minimum bias counter over threshold, was operated early in the data-taking period. 
It was used to trigger approximately $2\%$ of the integrated luminosity of the data sample analyzed. 
The lowest threshold jet trigger, uses a $0.4 \times 0.4$ window size in $\eta - \phi$ space at the first level of the trigger system 
and requires a jet with $p_T > 5~\giga\electronvolt/c$ at the “electromagnetic scale. Both triggers are found to be fully efficient for selecting events
with at least one jet with $p_T> 60~\giga\electronvolt/c$ and $|y|<2.8$.

The data correction for all trigger and detector efficiencies and resolutions, 
other than the energy scale correction already applied, is performed in a single step using a bin-by-bin migration 
function evaluated using validated MC samples. 
For each measured distribution, final-state-particle cross section using \texttt{Pythia MC09} is
evaluated in the relevant bins, along with the equivalent distributions as observed in the simulated detector.
The ratio of the distributions provides a correction factor which is then applied to the data as seen in the
detector.

The dominant source of systematic uncertainty is from the jet energy response. 
This scale uncertainty has been determined to be below 10\% over the whole kinematic range, 
and to be below 7\% for central jets with $p_T > 60~\giga\electronvolt$, neglecting possible topological correlations between jets.
This leads to an overall systematic uncertainty on the cross sections of around 40\% \cite{ATLAS-CONF-2010-050}.

% \paragraph{Theoretical models} 
The measured jet cross sections, corrected to particle-level, are then compared
to NLO perturbative QCD predictions.
The \texttt{NLOJET++} program \cite{ATLAS-CONF-2010-050} is used for all predictions, with \texttt{APPLgrid} \cite{ATLAS-CONF-2010-050} being used for
efficient evaluation of uncertainties.
Since the NLO calculations predict the value of partonic cross sections, which are not accessible,
they are corrected for the non-perturbative effects of hadronization and underlying event.
\texttt{Pythia MC09} is used to derive the baseline correction factors, and \texttt{Pythia Perugia0/Hard/Soft} and \texttt{Herwig+Jimmy} \cite{ATLAS-CONF-2010-050} 
are used for systematic variations. 
The main sources of theoretical uncertainties are the choice of renormalization 
and factorization scales, the parton distribution functions, $\alpha_s(M_Z)$
and the modeling of soft QCD effects (underlying event and parton hadronization processes) \cite{ATLAS-CONF-2010-050}.
The overall uncertainty is defined by the envelope of maximum variation around the prediction obtained with \texttt{Pythia MC09} \cite{ATLAS-CONF-2010-050}.

\section{RESULTS}
\begin{figure*}[htb]
\centering
%-------------------------------------
\subfigure%[Caption of subfigure 1]
{
\includegraphics[width=0.37\textwidth]{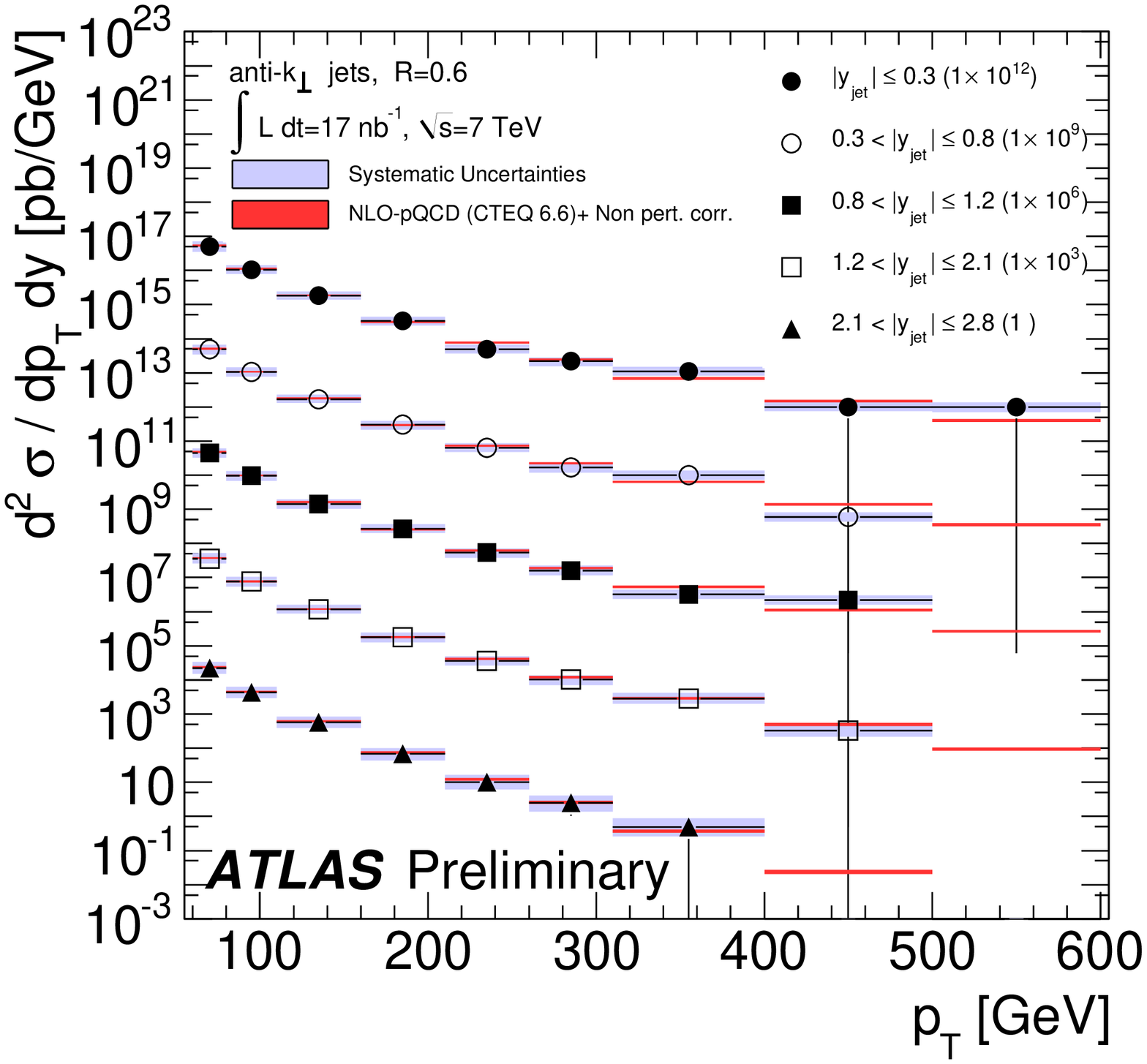}
\label{subfig:fig1}
}
%-------------------------------------
 \vspace{-0.4cm}
\hspace{1 cm}
%-------------------------------------
\subfigure%[Caption of subfigure 1]
{
\includegraphics[width=0.37\textwidth]{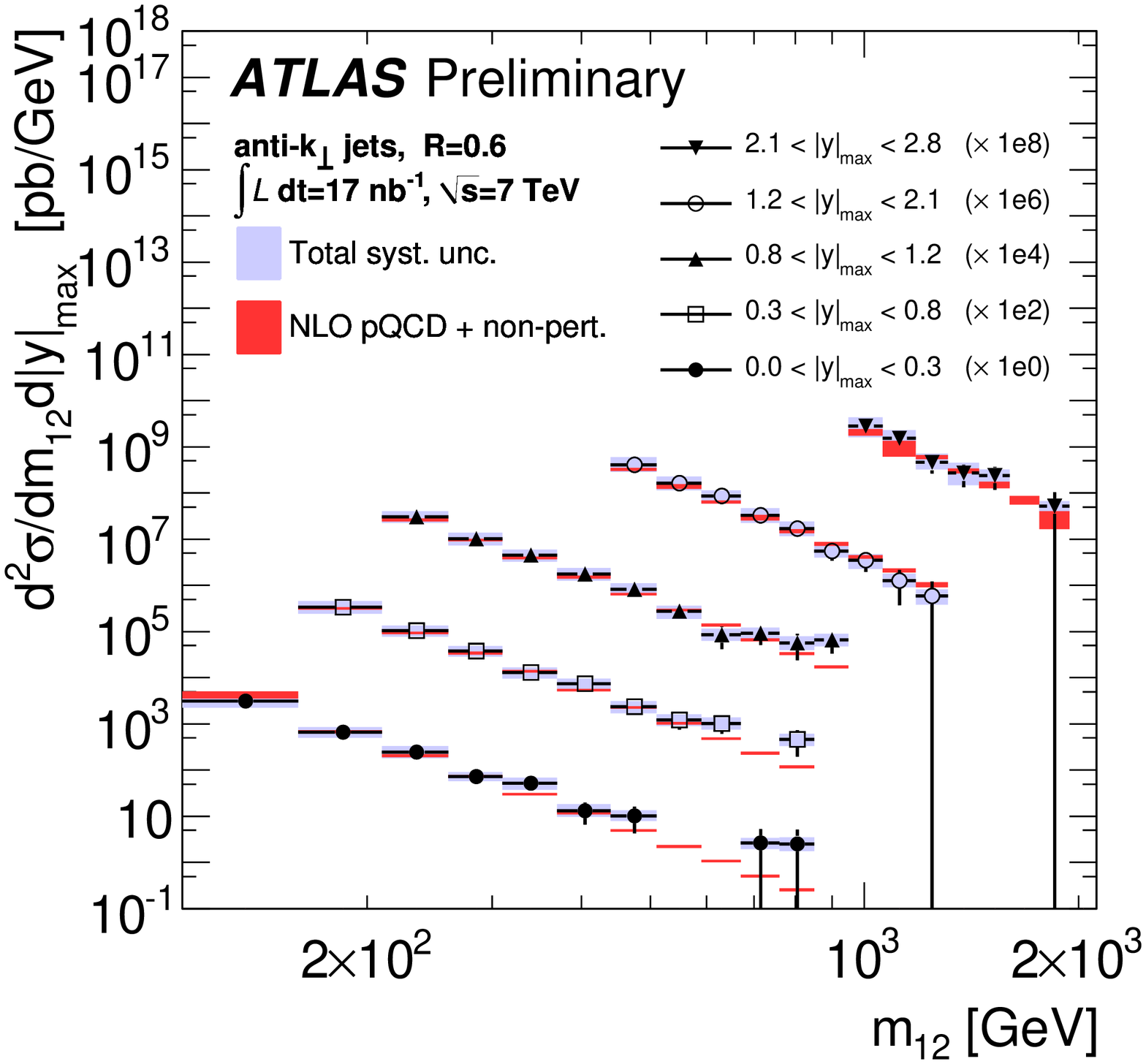}
\label{subfig:fig2}
}
%
%-------------------------------------
\subfigure%[Caption of subfigure 1]
{
\includegraphics[width=0.37\textwidth]{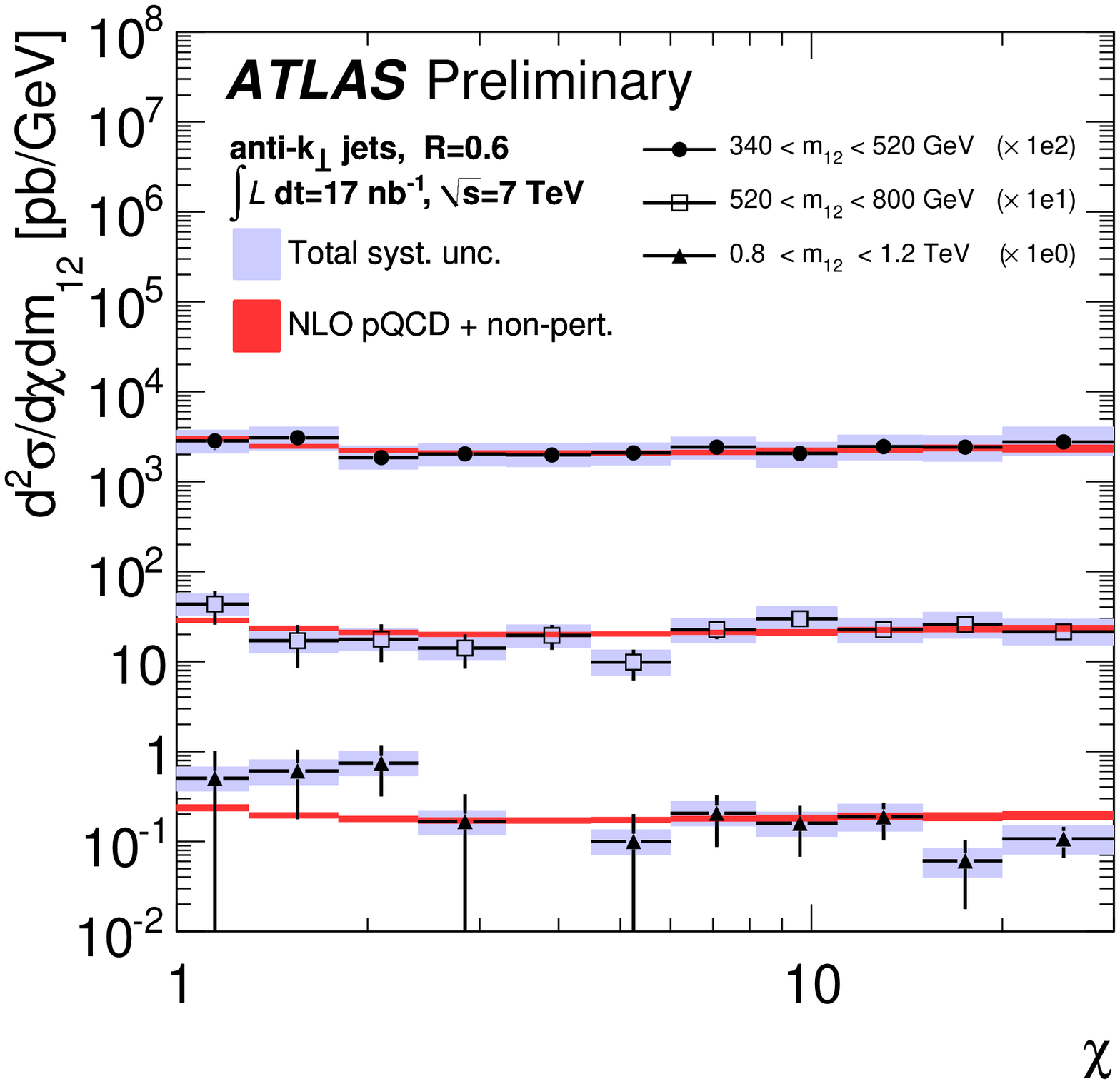}
\label{subfig:fig3}
}
%-------------------------------------
\hspace{1 cm}
%-------------------------------------
\subfigure%[Caption of subfigure 1]
{
\includegraphics[width=0.37\textwidth]{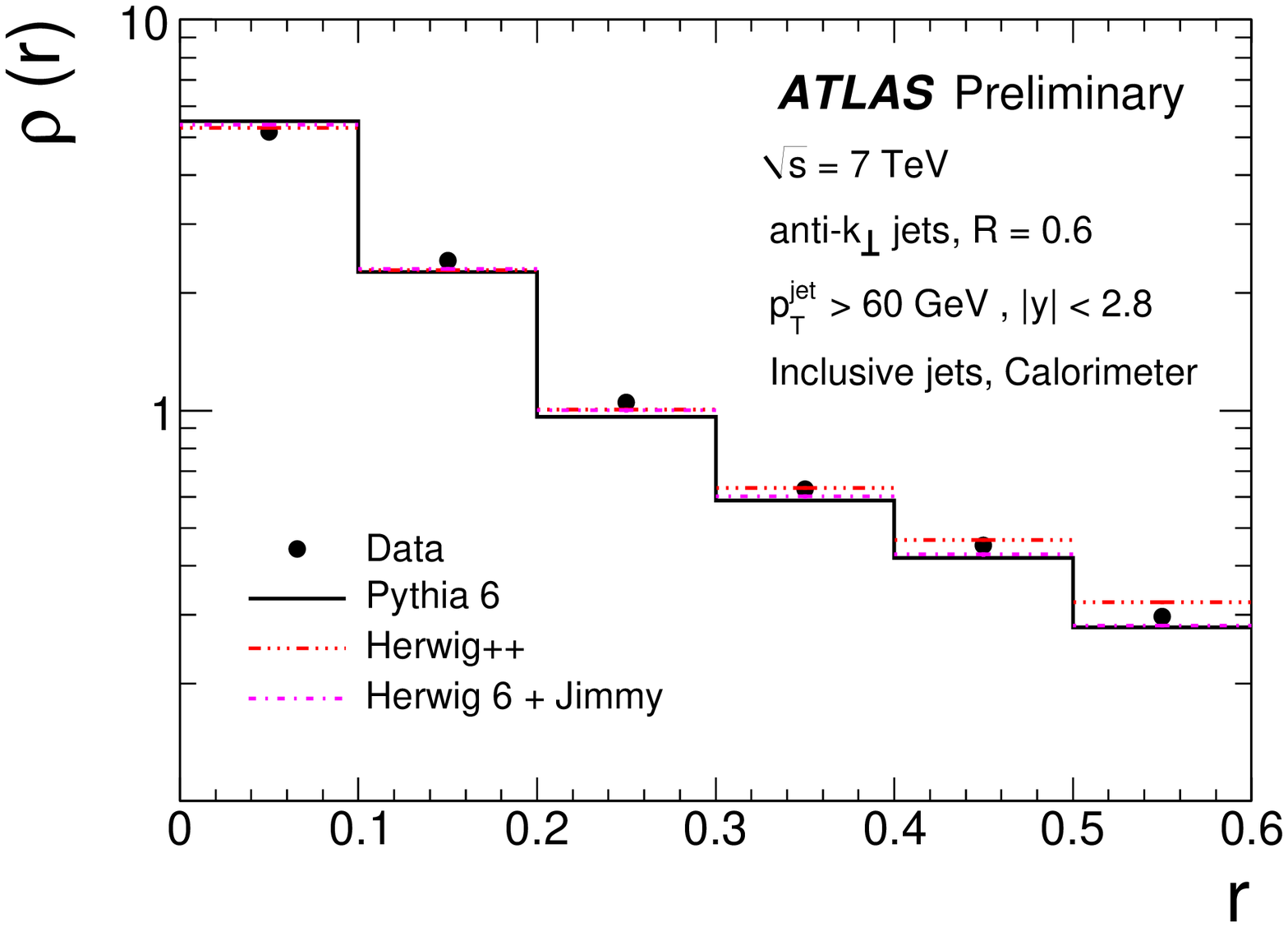}
\label{subfig:fig4}
}
%-------------------------------------
\caption{
\textbf{Top Left:} Inclusive jet double-differential cross section as a function of jet $p_T$ in different regions of $|y|$.
\textbf{Top Right:} Dijet double-differential cross section as a function of dijet mass, in the maximum $y$-bin between the two leading jets.
\textbf{Bottom Left:} Dijet double-differential cross section as a function of angular variable $\chi$ in different bins of the dijet mass.
\textbf{Bottom Right:} The uncorrected jet shape as a function of radius $r$, compared to MC simulations.
}
\label{fig:results}
\end{figure*}
Figure~\ref{subfig:fig1} shows the double-differential cross section as a function of $p_T$ in several different
regions of $y$. 
The cross section extends from $p_T = 60~\giga\electronvolt/c$ up to around $p_T = 600~\giga\electronvolt/c$, 
and falls by more than four orders of magnitude over this range.  
The error bars indicate the statistical uncertainty on the measurement, and the gray shaded band indicates 
the quadratic sum of the systematic uncertainties, dominated by the jet energy scale uncertainty.  
The theory uncertainty shown in orange is the quadratic sum of all uncertainties mentioned above.
In all $y$ regions, the theoretical predictions are consistent with the data.

In Figure~\ref{subfig:fig2} the double-differential dijet cross section is shown as a function of the dijet mass $m_{12}$,
for different bins of the maximum $y$ of the two leading jets.
The cross section falls rapidly with mass, but extends up to masses of nearly $2~\tera\electronvolt/c^2$. 
Figure~\ref{subfig:fig3} shows the cross section as a function of the dijet angular variable $\chi$ for different ranges of the dijet mass. 
Again, the theory is reasonably in agreement with the experimental results.
 
The jet shapes measured using calorimeter clusters, without applying any corrections for detector effects, are shown in Figure~\ref{subfig:fig4}. 
The jets simulated by \texttt{Pythia 6} are slightly narrower than the jets in the data, 
while the \texttt{Herwig 6 + Jimmy} and \texttt{Herwig++} simulations provide a better description.
\section{CONCLUSIONS}
Inclusive and dijet cross sections have been measured in $pp$ collisions at 
$\sqrt{s} = 7~\tera\electronvolt$ using the first data collected by the ATLAS detector. 
LO parton-shower MC programs provide a reasonable
description of the energy flow around the jets, and the cross section distributions. 
The measurements are well described by NLO perturbative QCD calculations corrected for non-perturbative QCD
effects.
\begin{acknowledgments}
This work is partlially supported by the European Commission, 
through the ARTEMIS Research Training Network, contract number MRTN-CT2006-035657.
\end{acknowledgments}

\end{document}